\documentclass[10pt,conference]{IEEEtran}
\IEEEoverridecommandlockouts

\usepackage{cite}
\usepackage{amsmath,amssymb,amsfonts}
\usepackage[bottom=4.2cm,top=2cm,left=1.57cm,right=1.57cm,columnsep=0.21in]{geometry}
\usepackage{algorithmic}
\usepackage{graphicx}
\usepackage{epstopdf}
\usepackage{algorithm}
\usepackage{textcomp}
\usepackage{xcolor}
\usepackage{amsthm}
\usepackage[utf8]{inputenc} 
\usepackage[T1]{fontenc}    
\usepackage{float} 
\usepackage{subcaption}

\newtheorem{remark}{Remark}
\def\BibTeX{{\rm B\kern-.05em{\sc i\kern-.025em b}\kern-.08em
    T\kern-.1667em\lower.7ex\hbox{E}\kern-.125emX}}
\begin{document}
\title{\fontsize{21}{12}\selectfont On the Multi-Weight Shift-Sum Iterative Decoding of Short BCH Codes}
\author{\IEEEauthorblockN{Yinglei Yang\IEEEauthorrefmark{1}, Di Zhang\IEEEauthorrefmark{2}, Aimin Li\IEEEauthorrefmark{3}, Yifei Shen\IEEEauthorrefmark{4}, Alexey Frolov\IEEEauthorrefmark{5}, Bo Gu\IEEEauthorrefmark{2}}
\IEEEauthorblockA{\IEEEauthorrefmark{1}School of Electrical and Information Engineering,
Zhengzhou University, Zhengzhou 450001, China}
\IEEEauthorblockA{\IEEEauthorrefmark{2}School of Intelligent Systems Engineering, Sun Yat-sen University, Shenzhen 518107  \& Southern Marine Science and \\ Engineering Guangdong Laboratory (Zhuhai), Zhuhai 519082, China}
\IEEEauthorblockA{\IEEEauthorrefmark{3} Department of Electronic Engineering, Middle East Technical University, Ankara 06800, Turkiye}
\IEEEauthorblockA{\IEEEauthorrefmark{4} Telecommunications Circuits Laboratory, \'{E}cole Polytechnique F\'{e}d\'{e}rale de Lausanne (EPFL), Lausanne 1015, Switzerland}
\IEEEauthorblockA{\IEEEauthorrefmark{5} 6G Technologies Laboratory, Skolkovo Institute of Science and Technology (Skoltech), Moscow, Russia}}



\maketitle

\begin{abstract}
This paper investigates a shift-sum decoding algorithm that employs dual codewords.  Existing dual codeword shift-sum decoding schemes mainly rely on minimum-weight dual codewords. However, the number of such codewords is often limited, which restricts the amount of reliability information available to the decoder. To address this issue, we extend the component counting mechanism from minimum-weight dual codewords to dual codewords whose Hamming weights lie within a prescribed upper bound. We further introduce a normalized component counting rule to suppress the overcounting bias caused by the much larger populations of higher-weight classes. Based on this framework, we develop a hard-decision iterative error reduction decoder and a soft information-aided iterative error reduction decoder. Numerical results show that the proposed method consistently outperforms existing minimum-weight shift-sum baselines especially for short codes where additional dual codeword information is particularly valuable.
\end{abstract}


\section{Introduction}
As classic linear cyclic block codes, Bose-Chaudhuri-Hocquenghem (BCH) \cite{bose1960class} codes serve as a cornerstone of channel coding theory, and their encoding and decoding algorithms have long been a core research focus in communication theory \cite{Xu2025}. BCH codes have been widely applied in high reliability scenarios of both digital communications and data storage \cite{Zhang2024Tii}, owing to their well defined algebraic structure and precisely tunable error correcting capability \cite{Zheng2025}. 
Among classic algorithms, Berlekamp-Massey (BM) algorithm achieves hard-decision decoding by iteratively solving syndrome equations, making it the most classic fundamental scheme in engineering applications \cite{Massey1969, Daniel2025,Sukmadji2024,Yu2016}. In scenarios where decoders can receive soft information, such as future wireless and optical communications, soft-decision decoding  significantly improves error correcting performance compared to hard-decision decoding \cite{Chen2024}. The Chase algorithm and Ordered Statistical Decoding (OSD) generate multiple test vectors for verification using soft information, which significantly improves decoding performance but at the cost of increased computational complexity \cite{Fossorier1995, Chase1972, Yang2025}. The Belief Propagation (BP) algorithm iteratively updates bit probability information based on the Tanner graph, exhibiting excellent performance \cite{Shen2025}. In recent years, a series of new decoding schemes have emerged continuously, such as neural network decoding \cite{Nadeem2026}, Guessing Random Additive Noise Decoding (GRAND) \cite{Duffy2022} and shift-sum decoding based on minimum weight dual codewords \cite{bossert2022hard}.

Shift-sum decoding is particularly attractive because it can be implemented by finite-field polynomial multiplication and simple comparison operations, which makes it hardware friendly \cite{bossert2022hard, xing2023shift}. In early work, Bossert showed that dual codeword structure can support hard- and soft-decision decoding beyond half the minimum distance \cite{bossert1986hard}. In recent years, \cite{bossert2019iterative} designed an efficient decoding scheme for cyclic codes, which obtains reliability information through cyclic shifts and component counting to further improve decoding performance. Subsequently, \cite{bossert2022hard} further extended such decoding schemes to various code types, including BCH codes, Reed-Muller (RM) codes, and Reed-Solomon (RS) codes. Meanwhile, \cite{yuan2021plausibility,xing2023shift,xing2020iterative} proposed new shift-sum decoding methods suitable for non-binary cyclic codes based on the statistical distribution characteristics of frequency matrices.

However, most of the existing works rely only on minimum-weight dual codewords, and the number of such codewords is often too small to provide sufficiently rich reliability information \cite{xing2023shift}. In our prior work within the vector operation framework of linear block codes, we extended the decoding method relying solely on minimum-weight dual codewords to a decoding theory based on multi-weight dual codewords \cite{Zhang2025}. However, it cannot be directly adapted to cyclic codes constructed via polynomial operations. This paper addresses this limitation while preserving the original shift-sum decoding framework. Instead of restricting the decoder to minimum-weight dual codewords, we extend the component counting mechanism to dual codewords  within a prescribed weight range. This extension enables the acquisition of more reliability information, and thus boosts the decoding algorithm performance. At the same time, we introduce a normalized component counting rule so that the decoder benefits from additional dual codewords without being dominated by high-cardinality weight classes. Based on this flexible weight framework, we further introduce a hard-decision decoder, namely Iterative Error Reduction Decoding (IERD), and one soft information aided decoder, namely Soft Iterative Error Reduction Decoding (SIERD). Simulation results demonstrate that the performance of the proposed hard-decision decoding algorithm IERD is closer to the theoretical Hamming bound than that of the aforementioned hard-decision decoding algorithms, and its error correction capability can exceed the designed error correction capability $t$ of BCH codes. 

\section{Prerequisites}
The primary research object of this paper is the binary primitive BCH code, hereinafter referred to as BCH code for brevity. Given positive integers $m$ and $t$ satisfying $m \geq 3$ and $t < 2^m - 1$, there always exists a BCH$(n, k)$  code with block length $n = 2^m - 1$ and number of parity bits $n - k \leq mt$. The designed minimum distance of this code is $d_{\min} = 2t + 1$, enabling it to correct all bit errors of magnitude less than or equal to $t$ during transmission. For this reason, it is also termed a  BCH$(n, k, t)$ code.

Given $m$ and $t$, let $\alpha$ denote a primitive element of the finite field $\mathrm{GF}(2^m)$. For $0 < i \leq 2t$, let $\varphi_i(X)$ represent the minimal polynomial of $\alpha^i$, which can also be characterized as a  cyclotomic coset. The generator polynomial of the corresponding  BCH$(n, k, t)$ code is then given by
\begin{align}
\mathbf{g}(X) = \text{LCM} \left\{ \varphi_1(X), \varphi_2(X), \dots, \varphi_{2t}(X) \right\} \label{g(x)},
\end{align}
where $\text{LCM}$ denotes the least common multiple of polynomials. The generator polynomial $\mathbf{g}(x)$ is a product of irreducible polynomials. The parity-check polynomial of the BCH$(n, k, t)$ code is denoted by $\mathbf{h}(x)$, satisfying $\mathbf{g}(x)\mathbf{h}(x) = x^n - 1$. A corresponding codeword is expressed as $\mathbf{c}(x) = \mathbf{a}(x)\mathbf{g}(x)$, where $\mathbf{a}(x)$ is an information polynomial of degree less than $k$. 

The generator polynomial of the dual code of the $\text{BCH}(n, k, d)$ code is $\mathbf{h}(x) = (x^n - 1)/\mathbf{g}(x)$, which forms a $\text{BCH}(n, n - k, d^\bot)$ code. A codeword of the dual code is given by $\mathbf{c}^\bot(x) = \mathbf{a}(x)\mathbf{h}(x)$, where $\mathbf{a}(x)$ is an information polynomial of degree less than $n - k$. Therefore, the product of any codeword and any dual codeword equals zero, i.e.,
\begin{align}
\mathbf{c}(x)\mathbf{c}^\bot(x) = 0 \pmod{x^n - 1}. \label{dual}
\end{align}

It is shown  in \cite{1977theory}  that for a BCH$(n, k, t)$ code, if it is capable of correcting $z$ errors (which requires satisfying $d_{\min} \geq 2z + 1$), then the following inequality holds
\begin{align}
M\cdot\sum_{i=0}^{z}\binom{n}{i}\leq 2^n,
\end{align}
where $M$ denotes the maximum number of codewords in the codebook with $M\leq2^k$, and $2^n$ is the total number of binary vectors of length $n$ (i.e., all possible received vectors). This inequality is also referred to as the Hamming Bound in coding theory. Thus, the maximum number of errors correctable by the hard-decision decoding scheme for a BCH$(n, k, t)$ code is given by:
\begin{align}
\tau_{\mathrm{max}} =\max\left\{ z \in \mathbb{N} \mid \sum_{i=0}^{z}\binom{n}{i} \leq 2^{n-k} \right\}.
\label{HB}
\end{align}

\section{Decoding Based On The Dual Codewords}

At the receiving end, we assume the received vector is $\mathbf{r}\left( x\right)  = \mathbf{c}\left( x\right)  + \mathbf{e}\left( x\right)$, where $\mathbf{e}\left( x\right)  = {x}^{{e}_{0}} + {x}^{{e}_{1}} + \ldots  + {x}^{{e}_{\tau  - 1}}$ is an error polynomial with weight $\tau$, and $\mathbf{c}\left( x\right)$ is the transmitted codeword. We can calculate
\begin{align}
\mathbf{r}\left( x\right) {\mathbf{c}}^{ \perp  }\left( x\right)  &= \left( {\mathbf{c}\left( x\right)  + \mathbf{e}\left( x\right) }\right) {\mathbf{c}}^{ \perp  }\left( x\right) \notag\\ 
&= \mathbf{e}\left( x\right) {\mathbf{c}}^{ \perp  }\left( x\right) \mod \left( {{x}^{n} - 1}\right).
\end{align}

Since the dual code is cyclic, we can select a dual codeword $ \mathbf{b}\left( x\right)$ with weight $\delta$ in the form of $\mathbf{b}\left( x\right)  = 1 + {x}^{{b}_{1}}+\ldots+ {x}^{{b}_{{\delta }^{ \perp  } - 1}}$. According to \eqref{dual}, we compute $\mathbf{w}\left( x\right)  = \mathbf{r}\left( x\right) \mathbf{b}\left( x\right) \mod \left( {{x}^{n} - 1}\right)$, which equals $\mathbf{w}\left( x\right)  = \mathbf{e}\left( x\right) \mathbf{b}\left( x\right) \mod \left( {{x}^{n} - 1}\right)$. The polynomial $\mathbf{w}\left( x\right)$ can be expressed as the sum of cyclic shifts of $\mathbf{e}\left( x\right)$ \cite{bossert2022hard}, i.e.,
\begin{align}
\mathbf{w}\left( x\right)  &= {x}^{{e}_{0}} + {x}^{{e}_{1}} + \ldots  + {x}^{{e}_{\tau  - 1}} + \notag\\
&{x}^{{e}_{0} + {b}_{1}} + {x}^{{e}_{1} + {b}_{1}} + \ldots  + {x}^{{e}_{\tau  - 1} + {b}_{1}} + \notag\\
&\quad \quad\quad\quad\quad\quad\quad \vdots \notag\\
&{x}^{{e}_{0} + {b}_{{\delta }^{ \perp  } - 1}} + {x}^{{e}_{1} + {b}_{{\delta }^{ \perp  } - 1}} + \ldots  + {x}^{{e}_{\tau  - 1} + {b}_{{\delta }^{ \perp  } - 1}},
\end{align}
where the exponents ${e}_{i} + {b}_{j}$ are calculated modulo $n$. Selecting a dual codewords  with a coefficient of $1$ for the zero-degree term ensures that, in the absence of shift errors, the coefficient of $\mathbf{w}\left( x\right)$ corresponding to the error position is $1$. This makes the probability that the coefficient corresponding to an error position is $1$ greater than that of a non-error position. This forms the basis of our subsequent decoding algorithm. Next, we will sequentially present the probability that the coefficient of $\mathbf{w}\left( x\right)$ at a non-error position is $1$ and the probability that the coefficient of $\mathbf{w}\left( x\right)$ at an error position is $1$.

Different from \cite{xing2023shift}, which derives the corresponding probabilities from the distribution of error positions for a fixed minimum dual codeword weight, we formulate the analysis from the possible non-zero positions of a generic dual codeword with weight $\delta$. This formulation accommodates dual codewords of multiple weights. For a position where the coefficient of the error polynomial is 0, to make the corresponding coefficient of $\mathbf{w}\left( x\right)$ at this position~$1$ (e.g., $x^i$), we need an odd number of the $\delta-1$ positions corresponding to the non-zero degree coefficients of the dual codewords  overlap with the other $\tau$ non-zero positions of the error polynomial such that ${x}^{{e}_{k} + {b}_{j}} \mod \left( {{x}^{n} - 1}\right) = x^i$. 
We assume that only $k$ non-zero positions in the error polynomial overlap with the $\delta-1$ positions corresponding to the non-zero degree coefficients of the dual codeword with weight~$\delta$. The corresponding probability can be expressed as the ratio of the product of the number of combinations of choosing $k$ positions from $\tau$ positions and the number of combinations of choosing $\delta-1-k$ positions from $n-\tau-1$ positions to the number of combinations of directly choosing $\delta-1$ positions from $n-1$ positions, as shown in the following formula
\begin{align}
Pr(k,\tau,\delta) =\frac{\binom{\tau}{k} \binom{n-\tau-1}{\delta-1-k}}{\binom{n-1}{\delta-1}}.
\end{align}

Thus, the expected probability that the coefficient of $\mathbf{w}\left( x\right)$ at a non-error position is $1$ is
\begin{align}
W(\delta-1,\tau) = \sum_{\substack{k=1 \\ k \text{ is odd}}}^{\tau} \frac{\binom{\tau}{k} \binom{n-\tau-1}{\delta-1-k}}{\binom{n-1}{\delta-1}}.
\label{W8}
\end{align}

For an error position (e.g., ${x}^{{e}_{i}}$), since the zero-degree coefficient of the selected dual codewords  is fixed to 1, we can stably obtain a coefficient of $1$ for ${x}^{{e}_{i}}$ without cyclic shifting.  Thus, to make the coefficient of $\mathbf{w}\left( x\right)$ at this position 1, it is required that an even number of combinations of the other $\tau-1$ non-zero positions of the error polynomial and the other $\delta-1$ positions of the dual codewords are shifted to ${x}^{{e}_{i}}$. The corresponding probability is
\begin{align}
W_e(\delta,\tau) &= 1- W(\delta-1,\tau-1) \notag\\& = 1- \sum_{\substack{k=1 \\ k \text{ odd}}}^{\tau-1} \frac{\binom{\tau-1}{k} \binom{n-\tau}{\delta-1-k}}{\binom{n-1}{\delta-1}}.
\label{W9}
\end{align}


\begin{figure}[!t]
\centering
\includegraphics[width=2.9in]{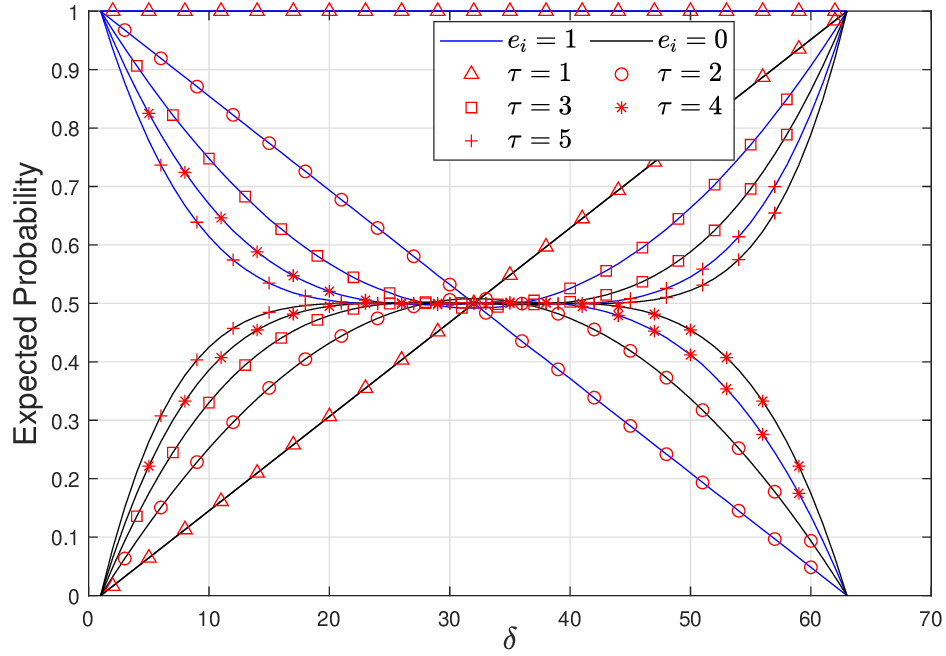} 
\caption{Expected probability in BCH$(n = 63)$.}
\label{fig1}
\end{figure}

For the idea of shift-sum decoding based on dual codewords, the most important thing to obtain effective key reliability information is that the probability that the coefficient corresponding to an error position is $1$ is greater than that of a non-error position. However, this point may not have been effectively proven for cyclic codes. Therefore, we need to prove the following equation holds
\begin{align}
&1- W(\delta,\tau-1) \geq W(\delta,\tau) \notag\\
&=1- \sum_{\substack{k=1 \\ k \text{ odd}}}^{\tau-1} \frac{\binom{\tau-1}{k} \binom{n-\tau}{\delta-1-k}}{\binom{n-1}{\delta-1}} \geq  \sum_{\substack{k=1 \\ k \text{ odd}}}^{\tau} \frac{\binom{\tau}{k} \binom{n-\tau-1}{\delta-1-k}}{\binom{n-1}{\delta-1}}.
\end{align}
Next, based on our prior work \cite[Theorem 1]{Zhang2025}, 
under the condition $2\tau + 2 + (\sqrt{\tau}+1)(\delta-3) \leq n$, we have
\begin{align}
1 - W(\delta,\tau) \geq W(\delta,\tau).
\end{align}

Moreover, $W(\delta,\tau)$ is monotonically increasing with respect to $\tau$. Therefore
\begin{align}
1- W(\delta-1,\tau-1) \geq 1- W(\delta-1,\tau) \geq W(\delta-1,\tau).
\end{align}
Thus, under $2\tau +3 +(\sqrt{\tau}+1)(\delta-4) \leq n$ conditions, $1- W(\delta-1,\tau-1) \geq W(\delta-1,\tau)$ also holds.

\begin{remark}[Consistency Between Theoretical Analysis and Simulation Results]
\label{Remark1}
As shown in Fig. \ref{fig1} for $n=63$, the theoretical probabilities in \eqref{W8} and \eqref{W9} agree well with the simulation results. More importantly, the foregoing analysis show that the distinction between error and non-error positions based on component counting is not restricted to minimum-weight dual codewords, but remains valid for dual codewords within the derived range of Hamming weights. This observation provides the theoretical basis for exploiting multi-weight dual codewords in the subsequent shift-sum decoding algorithms.
\end{remark}

Next, we present the specific method to calculate the component counting $\mathbf{\phi}$. Let $\mathbf{C} \subseteq  {\mathbb{F}}_{2}^{n}$ be a set of BCH codewords, and its dual codewords is defined as
\begin{align}
{\mathbf{C}}^{\perp}= \left\{\mathbf{b}(x) \mid {\forall }_{\mathbf{c}(x) \in \mathbf{C}} \mathbf{b}(x)\mathbf{c}(x)=0 \mod (x^n-1) \right\}.\notag
\end{align}
 Within the dual codewords  ${\mathbf{C}}^{ \perp  }$, we define the set
\begin{align}
\mathbf{B} &= \left\{  {\mathbf{b}(x) \in  {\mathbf{C}}^{ \perp  }\mid {wt}\left( \mathbf{b}(x)\right) < {d}_{B} \leq  \frac{n - {2\tau } + 4\sqrt{\tau } + 1}{\sqrt{\tau } + 1}}\right\}   \notag\\&= \left\{  {{\mathbf{b}}_{1}(x),\cdots ,{\mathbf{b}}_{s}(x)}\right\}   \subseteq  {\mathbf{C}}^{ \perp  }, 
\label{set_B}
\end{align}
where $\left| \mathbf{B}\right|$ denotes the number of elements in $\mathbf{B}$, and the elements of set $\mathbf{B}$ are also referred to as decoding vectors.

Given the received codeword $\mathbf{r}(x) = \mathbf{c}(x) + \mathbf{e}(x)$, where $\mathbf{c}(x) \in  \mathbf{C}$ and $\mathbf{e}(x)$ is the error polynomial, the calculation of component counting $\mathbf{\phi}$ can be expressed as
\begin{align}
\mathbf{\phi}  &= \sum _{{\mathbf{b}(x) \in  \mathbf{B}}}\mathbf{b}(x)\mathbf{r}(x) \mod (x^n-1)  \notag\\&= \sum _{{\mathbf{b}(x) \in  \mathbf{B}}}\mathbf{b}(x)\mathbf{e}(x) \mod (x^n-1),
\end{align}
where $\mathbf{\phi}$ is approximately equal to the value in the following formula. The larger $\left| \mathbf{B}\right|$ is, the higher the consistency between the calculated value $\mathbf{\phi}$ and the theoretical value. 

\begin{align}
&\mathbf{\phi}_i  \\
&\approx\begin{cases}
\sum \limits_{{\mathbf{b}(x) \in  \mathbf{B}}}(W_e\left( {{wt}\left( \mathbf{b}(x)\right), {wt}\left( \mathbf{e}(x)\right) }\right)) & \text{if } \mathbf{e}_i(x) =1, \\
\sum \limits_{{\mathbf{b}(x) \in  \mathbf{B}}}W\left( {{wt}\left( \mathbf{b}(x)\right)-1, {wt}\left( \mathbf{e}(x)\right) }\right) & \text{if } \mathbf{e}_i(x) =0. \notag
\end{cases}
\end{align}
where \(e_i(x)\) represents the coefficient of \(x^{i-1}\) in \(e(x)\). Since for any dual codeword $\mathbf{b}(x) \in \mathbf{B}$, the probability that the coefficient corresponding to an error position is $1$ is greater than that of a non-error position, the position with a larger coefficient in $\mathbf{\phi}$ has a higher probability of being an error position.

\section{Decoding algorithm}
In our our proposed schemes, the first step is to randomly generate the set $\mathbf{B}$ of relevant dual codewords with weight bounded above by $d_B$, where the value of $d_B$ and $ \left| \mathbf{B}\right|$  are determined by the code length and the desired decoding performance. At the receiver, we compute $\mathbf{\phi}$ and sort the obtained $\mathbf{\phi}$ values according to their reliability. Since the dual codewords adopted herein have multiple weights, the number of dual codewords with larger weights may be far greater than that of dual codewords with the minimum weight. Therefore, dual codewords with larger weights may exert a more significant influence on the decoding process. However, as analyzed earlier, the information provided by dual codewords with smaller weights is more effective in distinguishing error positions from non-error positions. To mitigate the degradation of decoding performance caused by the disparity in codeword counts, we introduce a modification to the calculation of $\mathbf{\phi}$ here. We use $\mathbf{wn}$ to record the number of dual codewords for each distinct weight, where the element value corresponding to $\mathbf{wn}_i$ denotes the number of dual codewords with weight $i$. Thus, $\mathbf{\phi}$ can be computed as:
\begin{align}
\mathbf{\phi} &= \sum_{\mathbf{b}(x) \in \mathbf{B}} \frac{\mathbf{b}(x)\mathbf{r}(x) \pmod{(x^n-1)}}{\mathbf{wn}(wt(\mathbf{b}(x)))}  \notag \\ &= \sum_{\mathbf{b}(x) \in \mathbf{B}} \frac{\mathbf{b}(x)\mathbf{e}(x) \pmod{(x^n-1)}}{\mathbf{wn}(wt(\mathbf{b}(x)))}.
\label{phi}
\end{align}

The value of $\mathbf{\phi}$ obtained in this manner approximates the expression given below. The larger the non-zero element values in $ \mathbf{wn}$, the higher the consistency between the calculated value $\mathbf{\phi}$ and the theoretical value.
\begin{align}
&\mathbf{\phi}_i \approx   \notag\\
&\begin{cases}
\sum\limits_{\substack{i=1 \\\mathbf{wn}(i) \neq 0}}^{n} \left(1 - W_e\left(i, {wt}\left( \mathbf{e}(x)\right) \right)\right) & \text{if } \mathbf{e}_i(x) = 1, \\
\sum\limits_{\substack{i=1 \\\mathbf{wn}(i) \neq 0}}^{n} W\left( i-1, {wt}\left( \mathbf{e}(x)\right)\right) & \text{if } \mathbf{e}_i(x) = 0.
\end{cases}
\end{align}

Next, we propose two decoding schemes for BCH.

IERD Scheme: Iterative Error Reduction Decoding (IERD) is a hard-decision decoding method. In each iteration, the unreliable positions identified based on ${\mathbf{\phi}}$ are flipped to compute a new received vector, as detailed in \textbf{Algorithm \ref{alg:AOA}}. Furthermore, the position corresponding to $\mathbf{\phi}_i = \max \left( \mathbf{\phi}\right)$ indicates the most probable error position where ${\mathbf{e}}_{i}\left( x\right) = 1$. Based on this, we can iteratively reduce the number of errors by flipping these positions. When $\max \left( \mathbf{\phi}\right) > 0$, we update the received vector as $\mathbf{r} = \mathbf{r} + {\mathbf{e}}_{i}$ and continue the iteration. When $\max \left( \mathbf{\phi}\right) = 0$, it indicates that the correct transmitted codeword has been found. IERD can identify error positions in each iteration but requires recomputing $\mathbf{\phi}$ every time.
\begin{algorithm}[t!]
    \caption{The IERD Algorithm}
    \label{alg:AOA}
    \renewcommand{\algorithmicrequire}{\textbf{Input:}}
    \renewcommand{\algorithmicensure}{\textbf{Output:}}
    \begin{algorithmic}[1]
        \REQUIRE Received vector $\mathbf{r} = \mathbf{c}\left( x\right)  + \mathbf{e}\left( x\right)$,  max iterations \( T_{\mathrm{max}}\), and \(\mathbf{wn}\) 
        \ENSURE $\mathbf{c} $   
        \FOR{$k=1:T_{\mathrm{max}}$}
        \STATE calculate $\mathbf{\phi}$  \eqref{phi} 
        \STATE $j=\arg\max\limits_{i}\phi_{i}$ 
        \IF {$\mathbf{\phi}_j ==0$}
            \STATE return $\mathbf{r}$
        \ELSE
            \STATE $\boldsymbol{r} = \mathbf{r}+\mathbf{e}_j$
        \ENDIF
        \ENDFOR
    \end{algorithmic}
\end{algorithm}

Soft IERD (SIERD) Scheme: \textbf{Algorithm \ref{alg:AOA}} only utilizes the intrinsic information derived from dual codewords and fails to fully exploit the soft information available at the receiver. By incorporating soft information, for instance, in an Additive White Gaussian Noise (AWGN) channel, the received vector can be expressed as $\boldsymbol{r} = \boldsymbol{x} + \boldsymbol{n}$, where $x$ is the transmitted signal vector and $\boldsymbol{n}$ is the Gaussian noise vector with each component $n_i$ following a normal distribution with mean $0$ and variance $\sigma^2$. We note that a larger magnitude $|\boldsymbol{r}_i|$ of each component in the received vector corresponds to higher reliability of that position. This constitutes the soft information we can leverage. By fusing soft information with intrinsic information, we obtain the comprehensive information 
\begin{align}
{E}_{i} = LR \cdot \mathbf{\phi},
\end{align}
where $LR_i = \max{(|\boldsymbol{r}_i|)} - |\boldsymbol{r}_i|$. Similar to Algorithm 1, this comprehensive information is used to identify the most probable error positions where ${\mathbf{e}}_{i} = 1$. It is further employed to update the a priori information for the $i$-th position in the next iteration as $L{R}_{i} = \alpha {E}_{i}$, where $\alpha$ is a scaling factor. Compared with IERD, SIERD can significantly improve the accuracy of error position location, especially when the number of errors is large, thereby enhancing the decoding performance.

\section{Numerical Results}

For a given BCH$(n, k, t)$ code, we first determine the maximum number of potentially correctable errors $\tau_{\mathrm{max}}$ by calculating \eqref{HB}. We then determine the maximum weight of the required dual codewords $\delta_{\mathrm{max}}$ from \eqref{set_B}. Finally, a set of dual codewords $B$ is obtained through a random generation method. In all presented figures, higher index numbers appended to the proposed algorithm indicate an increased number of adopted dual codewords $\left| \mathbf{B}\right|$.

\begin{figure}[!t]
\centering
\includegraphics[width=2.9in]{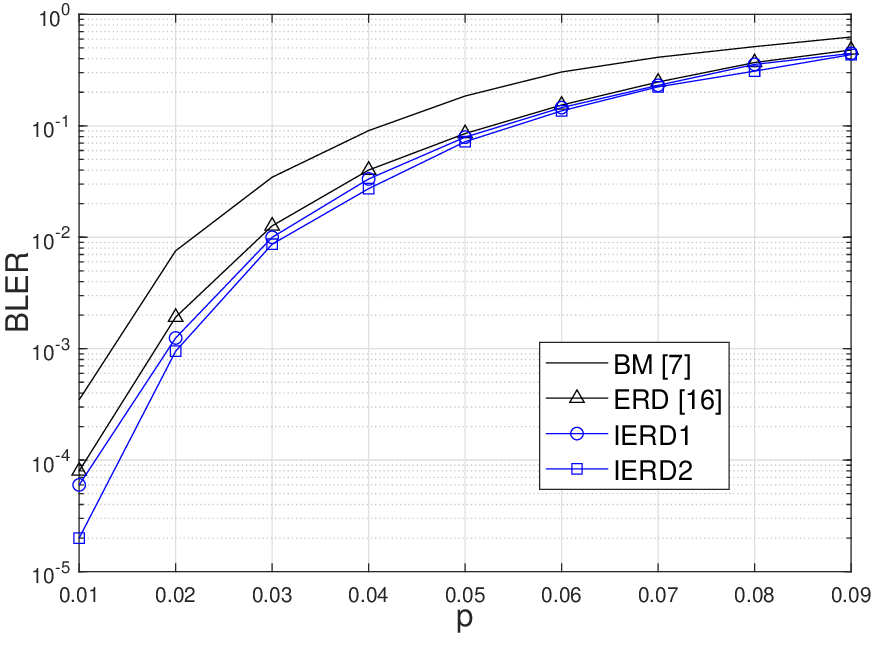} 
\caption{BLER performance of BCH$(63,39)$  over BSC channel.}
\label{fig2}
\end{figure}

\begin{figure}[!t]
\centering
\includegraphics[width=2.9in]{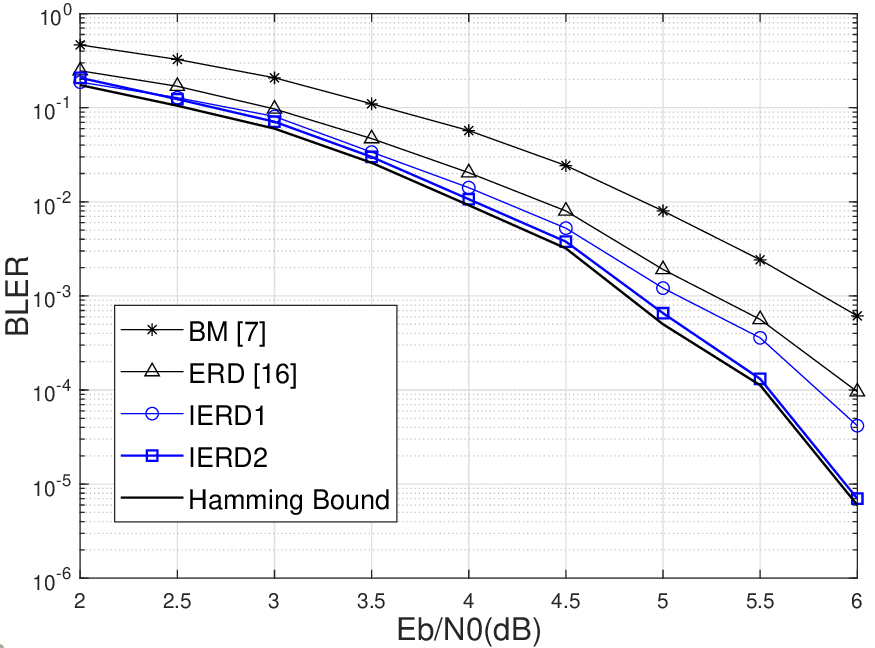} 
\caption{BLER performance of hard-decision decoding algorithms for the BCH$(63,30)$ code over an AWGN channel.}
\label{fig3}
\end{figure}
\begin{figure}[!t]
\centering
\includegraphics[width=2.9in]{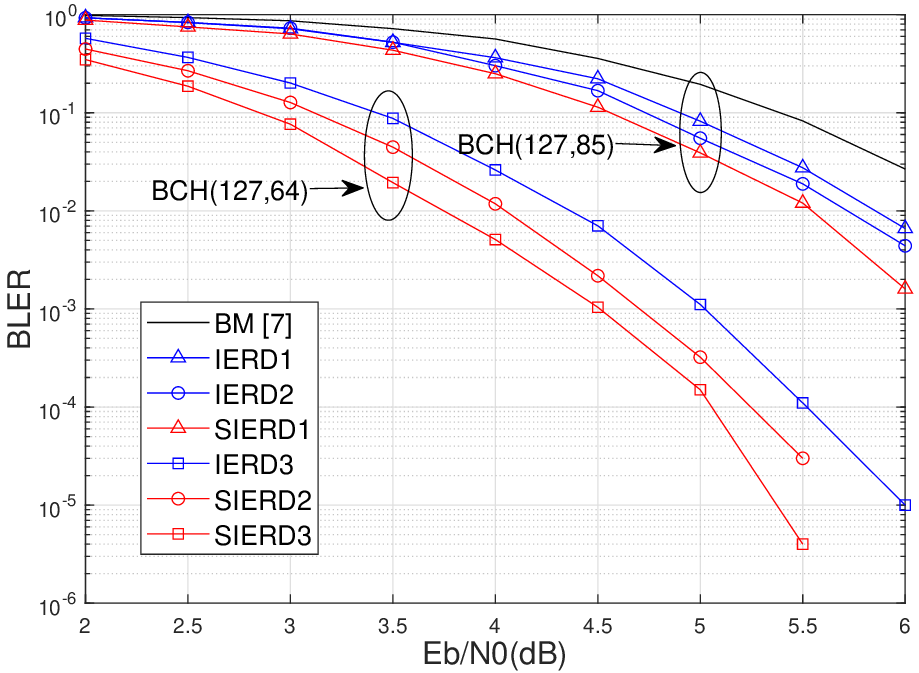} 
\caption{BLER performance of  BCH$(127, 64)$ and BCH$(127, 85)$ codes over AWGN channel.}
\label{fig4}
\end{figure}

Fig. \ref{fig2} depicts the Block Error Rate (BLER) performance of the proposed algorithm for the BCH$(63,39)$ code over the Binary Symmetric Channel (BSC), while Fig. \ref{fig3} analyzes the BLER performance of the algorithm for the BCH$(63,30)$ code over an AWGN channel. The results demonstrate that, with the same algorithm architecture, the scheme incorporating dual codewords of multiple weights achieves significantly superior performance compared with the one only utilizing dual codewords of the minimum weight (Error Reduction Decoding (ERD) in \cite{bossert2022hard}). In particular, it can be observed from Fig. \ref{fig3} that as the number of dual codewords involved in iteration increases, the performance curve of the proposed algorithm approaches the Hamming bound closely. This indicates that the algorithm can accurately correct up to $\tau_{\mathrm{max}}$ errors for the BCH$(63,30)$ code, whose error correcting capability exceeds the theoretical design limit of this code type. In addition, if the decoding process is assisted by the soft information of the channel, the algorithm can obtain better performance with fewer dual codewords, thereby reducing the computational complexity effectively. Fig. \ref{fig4} further verifies the effectiveness of the proposed algorithm for the BCH code with code length $n=127$ with different code rates. Although the increase in code length will increase the difficulty of searching for dual codewords with specific weights, the algorithm still maintains good applicability in the scenarios of medium and short code lengths, and it outperforms the classical BM algorithm. In terms of complexity, each iteration of the IERD scheme requires at most $|\mathbf{B}|\cdot\delta_{\mathrm{max}}^2 \cdot n$ finite field multiplications, and thus yields a worst-case computational complexity of $\mathcal{O}(T_{\mathrm{max}} \cdot|\mathbf{B}|\cdot\delta_{\mathrm{max}}^2 \cdot n)$. The SIERD scheme benefits from fewer iterations compared to the IERD scheme.  Furthermore, all the algorithms we designed can be implemented merely through polynomial multiplication and a small number of comparison operations, thus facilitating efficient hardware implementation in practical scenarios.

\section{Conclusion}
This paper studies multi-weight shift-sum decoding for short BCH codes. By extending the component counting analysis from minimum-weight dual codewords to dual codewords within a prescribed Hamming-weight range, we characterize the statistical distinction between error and non-error positions and enable additional reliability information to be exploited in decoding. A normalized component counting rule is further introduced to suppress the bias caused by high-cardinality weight classes, based on which the IERD and SIERD algorithms are developed. Numerical results show that IERD approaches the Hamming bound and can correct beyond the designed error correction capability of short BCH codes, while SIERD further improves decoding performance by incorporating channel soft information with fewer dual codewords.

\section*{Acknowledgement}
This work was supported by the Henan Natural Science Foundation for Excellent Young Scholar under Grant 242300421169, the Sun Yat-sen University Talent Initial Funds under Grant 76160-12256017 and 76160-12266015. 

\begin{bibliographystyle}{IEEEtran}
\begin{bibliography}{ref}
\end{bibliography}
\end{bibliographystyle}

\end{document}